\documentclass[10pt]{article}

\usepackage{natbib}
\usepackage{graphicx}
\usepackage{txfonts}
\usepackage[normalem]{ulem}
\usepackage{color}
\bibpunct{(}{)}{;}{a}{}{,}

\newcommand{\qm}[1]{``#1''}

\def\sss{\scriptscriptstyle}
\def\U{{\sss \!U}}
\def\L{{\sss \!L}}

\def\nuL{\nu_\L}
\def\nuU{\nu_\U}

\def\4u1636{\mbox{4U~1636$-$53}}
\def\Hz{\mathrm{Hz}}

\definecolor{gray}{rgb}{.6,.6,.6}
\definecolor{green}{rgb}{0,.6,0}
\definecolor{red}{rgb}{.9,0,0}

\begin{document}
\begin{center}
\noindent
{{\bf\Large On the origin of clustering of frequency ratios in the atoll source 4U~1636-53}}
\end{center}
\vspace{1ex}
\noindent
 {Gabriel T\"or\"ok$^{1}$
  ,~
  Marek A. Abramowicz$^{1,2,3}$
  ,~
  Pavel Bakala$^{1}$
  ,~
  Michal Bursa$^{4}$
  ,~
  Ji\v{r}\'{\i} Hor\'{a}k$^{4}$
  ,~
  Paola Rebusco$^{5}$
  ,~
  Zden\v{e}k Stuchl\'{\i}k$^{1}$
  }
\\\\
{\footnotesize{
  $^{1}$ Institute of Physics, Faculty of Philosophy and Science, Silesian
  University in Opava, Bezru\v{c}ovo n\'{a}m. 13, 746-01 Opava, CZ
  \\
  $^{2}$ Department of Physics, G\"oteborg University, S-412 96 G\"oteborg, SE
  \\
  $^{3}$ Copernicus Astronomical Centre PAN, Bartycka 18, 00-716 Warsaw, PL
 \\
  $^{4}$ Astronomical Institute of the Academy of Sciences, Bo\v{c}n\'{\i}~II 1401/1a, 141-31 Praha~4, CZ
 \\
  $^{5}$ MIT Kavli Institute for Astrophysics and Space Research, 77 Massachusetts Avenue, 37, Cambridge, MA 02139, US
\vspace{-3.5ex}

\noindent
{\scriptsize \\\\~~e-mail: terek@volny.cz, marek.abramowicz@physics.gu.se, pavel.bakala@ fpf.slu.cz, bursa@astro.cas.cz, \\horak@astro.cas.cz, pao@space.mit.edu, zdenek.stuchlik@fpf.slu.cz}  
}}
%
\\\\
{\footnotesize {\bf{Abstract.}}
A long discussion has been devoted to
the issue of clustering of the kHz QPO frequency ratios in the neutron star
sources. While the distribution of ratios inferred from an occurrence
of a single QPO seems to be consistent with a random walk, the distribution
based on simultaneous detections of both peaks indicates a preference
of ratios of small integers. Based on the public RXTE data we further investigate
this issue for the source \4u1636. Quality factors and rms amplitudes
of both the QPOs nearly equal to the points where the frequencies are
commensurable, and where the twin QPO detections cluster. We discuss a
connection of the clustering with the varying properties of the two
QPO modes. Assuming approximative relations for the observed correlations
of the QPO properties, we attempt to reproduce the frequency
and ratio distributions using a simple model of a random-walk evolution
along the observed frequency-frequency correlation. We obtain results
which are in qualitative agreement with the observed distributions.
}
\\\\{\footnotesize{{\bf Keywords:}{~X-rays: binaries --- Stars: neutron --- Accretion, accretion disks}}

\normalsize

\section{Introduction}

Since the paper of \cite{abr-etal:2003}, the issue of distribution of kHz~QPOs in neutron-star low mass X-ray binaries has been discussed extensively. In their work, \citeauthor{abr-etal:2003} examined simultaneous detections of the upper and lower QPOs in the Z-source Sco~X-1. The authors show that that the ratios of the lower and upper QPO frequencies cluster most often close to the value~$\nuL/\nuU\!=\!2/3$. They find also evidence for the second peak in a distribution of frequency ratios at $\nuL/\nuU\!\approx\!0.78$. This value is remarkably close to an another ratio of small 
integers, $4/5=0.8$. In the most recent paper, \citet{tor-etal:2007} have examined occurrences of the twin QPOs in the atoll source \4u1636\ applying the same methodology as \citet{abr-etal:2003}. They find that the distribution of the (inverse) frequency ratios $\nuU/\nuL$ of two simultaneously detected QPOs peaks near 3/2 and 5/4.

A preference of the commensurable frequency ratios in kHz QPO data of various sources has been systematically checked by a group of Belloni and his collaborators. \citet{bel-etal:2005} have re-examined the ratio distribution in Sco X-1 and later also in a larger sample comprising four atoll sources including \4u1636\ \citep{bel-etal:2005}. They argue that such clustering does not provide any useful information because frequencies of the two QPOs are correlated and the distribution of the ratio of two correlated quantities is completely determined by the distribution of one of them. Keeping this argument, a recent study of \citet{bel-etal:2007b} based on a systematic long term observation of \4u1636\ concludes that there is no preferred frequency ratio.

The aparent disagreement in conclusions of the two groups comes from a confusion between the observed frequency distribution (the one which can be recovered from observed data) and the intrinsic distribution (the ``invisible'' one really produced by the source). While \citet{abr-etal:2003} and \citet{tor-etal:2007}  have examined frequency ratios of the actually observed QPO pairs (twin peaks) only, the analysis of~\citet{bel-etal:2005,bel-etal:2007b} study primarily distributions of frequencies of a single QPO and make implications for the distribution of the other, often invisible, QPO from the empirical correlation between frequencies.

In this paper we show that the observed distributions are affected by the way the signal form a source is being detected and analyzed. We show that the observed clustering can be understood in terms of rms amplitude and quality factor correlations with QPO frequency. Taking these correlations into account, we simulate the ratio distribution using a random walk model of QPO frequency evolution and we find that results of the simulation agree with empirical data.


\section{Properties of oscillation modes on large frequency range}

{In the process of data reduction and searching for QPOs, an important quantity is the significance~$S$ of the peak in PDS, which measures the peak prominence. Shape of a peak in the PDS is most often fitted by a Lorentzian. Usually, \mbox{$S\geq2\!-\!4$} is being used as the low threshold limit for detections and only peaks that have their significances greater than tis limit are considered as QPOs. Thus, this imposes a certain selection criterion, which could consequently affect the distribution of detections.}

The significance $S$ is given by the relation between the integral area of a Lorentzian in PDS and its error. For a particular detection, it depends on observational conditions, on the quality factor $Q$ of the peak (defined as the QPO centroid frequency over the peak full-width at its half-maximum) and on the fractional root-mean-squared amplitude $r$ (a measure for the signal amplitude given as a fraction of the total source flux that is proportional to the root mean square of the peak power contribution to the total power spectrum), \mbox{$S = k\,r^2\!\sqrt{Q/\nu}$}, where the time-varying factor \mbox{$k(t) = I(t) \sqrt{T}$} depeds on the total length of observation~$T$ and the instantneous source intensity~$I$, which at a given time same for both upper and lower peak\footnote{The standard process of the QPO determining is in detail described in \cite{kli:1989}.}.

\cite{bar-etal:2005a,bar-etal:2005b,bar-etal:2005c,bar-etal:2006} 
have shown that both quality factors and rms amplitudes are determined by frequency and moreover that their profiles greatly differ between lower and upper QPO modes. The quality factor of the upper QPO is usually small and tends to stay at an almost constant level around \mbox{$Q_\U\sim10$}. In contrast, the lower QPO quality factor improves with frequency and can reach up to \mbox{$Q_\L\sim200$} before a sharp drop of coherence at high frequencies. Amplitudes of upper QPOs generally decrease with frequency, while the lower QPO amplitudes show first an increase and then they start to decay too.

Figure~\ref{fig:Q_rms_sig} shows the behaviour of amplitudes and quality factors of individual QPO modes in \4u1636\ and how they change with frequencies.\footnote{In the figure we use a correlation \mbox{$\nuU=0.701\nuL+520$Hz} from \citeauthor{abr-etal:2005} 2005 (see also \citeauthor{bel-etal:2005} 2005; \citeauthor{zha-etal:2006} 2006).} The displayed data of \cite{bar-etal:2005b} cover large frequency range available through the shift-add technique over all RXTE observations \citep[see][for details]{men-etal:1998,men-etal:1999,bar-etal:2005a,bar-etal:2005b,bar-etal:2005c}.
 
\begin{figure*}[t]
\includegraphics[width=\textwidth]{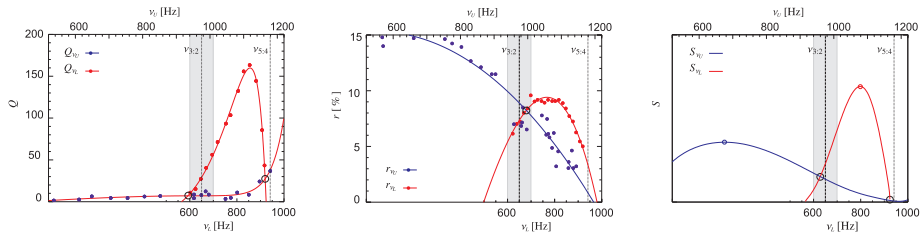}
\caption{\footnotesize The quality factor ({\sl left}), rms amplitude ({\sl middle}) and inferred significance ({\sl right}) behaviour in atoll source \4u1636. Red points represent lower QPO data, blue points are for upper QPO data. Data in first two panels comes from the study of \cite{bar-etal:2005b} and cover large range of frequencies available via shift-add method through all segments of RXTE observations. Continuous curves are obtained from interpolation by several exponentials \citep[see, e.g.,][]{tor:2007}. The prospected course of the QPO significance in the right panel is determined by the rms amplitude and quality factor profiles ($S\!\propto\!r^2\sqrt{Q/\nu}$). Frequency axes are related using frequency correlation ($\nuU=0.701\nuL+520$Hz; \citeauthor{abr-etal:2005} 2005).}
\label{fig:Q_rms_sig}
\end{figure*}

Note that both of the two properties are becoming similar as the frequency approaches points corresponding to $3/2$ or $5/4$ ratio \citep[the equality of amplitudes have been reported by][]{tor:2007}. In the right panel of Figure~\ref{fig:Q_rms_sig} we then plot the significances of the two oscillation modes, inferred from the combination on the two plots, while we keep the intensity $I$ and observing time $t$ constant (for simplicity). It is clearly visible that there is a similar equality of QPO significances close to points, where the frequencies are close to the $3/2$ or $5/4$ ratio (as a result of comparable $Q$ and $r$ at those points), while they are much different elsewhere. We will hereafter call the points of equal significances as the ``$3/2$'' and ``$5/4$'' points. We may also observe that the upper QPO mode is usually strong (much more significant) left from the $3/2$ point (at lower frequencies), while right from $3/2$ the lower QPO mode dominates.

\section{Clustering of frequency ratios}


It is likely that if QPOs are produced in a source, they are always produced in pairs. Because the strength of oscillations is usually around the sensitivity threshold of measurements, often only one (the stronger) QPO is detected. Around the special points $3/2$ and $5/4$, where significances are comparable, there is a good chance that if one mode can be detected the other could be detected as well, because both peaks have nearly the same properties. Indeed, this agrees with what is observed and has been laboured or challenged many times \citep{abr-etal:2003,bel-etal:2005,bulik:2005,yz:2007,bel-etal:2007b} that pairs of QPOs cluster close to the $3/2$ and some other small rational number ratios.

From time to time, the conditions at the source become such that both QPOs can be detected simultaneously regardless of their frequency, only because of their actual high brightness (as the observational sensitivity is relatively low). These events allows us not only to see QPO pairs close to the critical points, but sporadically also all the way along the frequency-frequency correlation line, even far from $3/2$. 

The clustering of frequency ratios close to $3/2$ is in this view significantly affected by the behaviour of rms amplitudes and quality factors and namely by the fact that these quantities become equal close to that frequency ratio. This is demonstrated in Figure~\ref{fig:simulation} (left), where we show a fraction of number of observations, in which both QPOs have been detected simultaneously, to a number of those, in which at least one QPO has been detected. The figure is based on data used in \cite{tor-etal:2007}. Clearly, the positions of maxima remarkably well correlate with points, where the two significances equal. Moreover, these positions coincide with peaks in the distribution of frequency ratios found in \cite{tor-etal:2007} which justify a hypothesis that there is a link between QPO properties and the ratio clustering.

\begin{figure*}[t]
\includegraphics[width=1\textwidth]{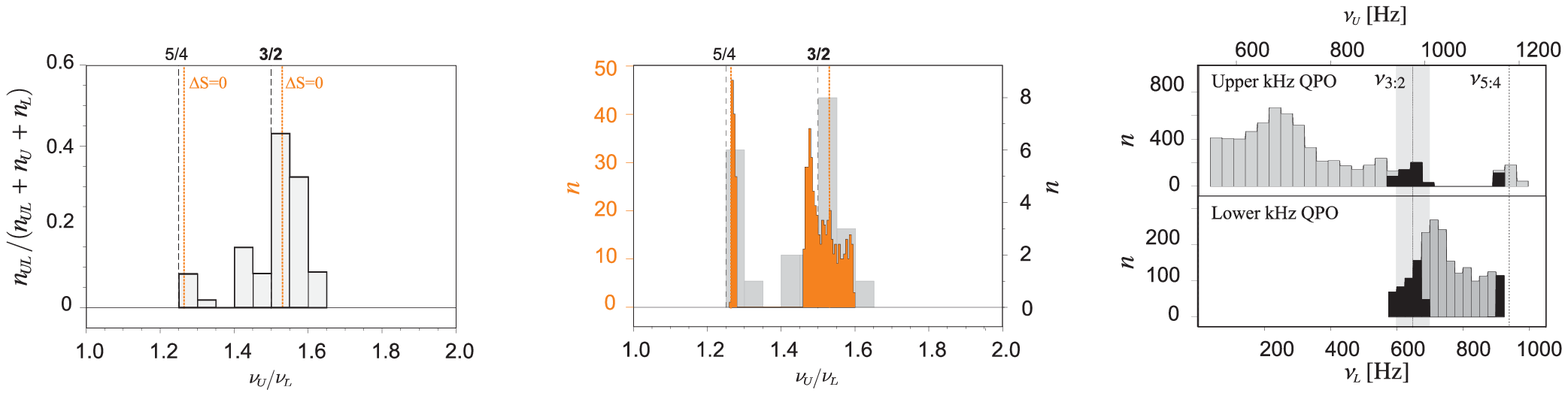}
\caption{\footnotesize  The distribution of observed frequency ratios. {\sl Left:} The fraction of the number of observations with simultaneous detections $n_{\U\L}$ to the number of observetions in which at least one QPO has been detected $n_{\U\L} + n_\U + n_\L$ (where $n_\U$ and $n_\L$ are respectively the numbers of observations with detections of the upper or lower QPO only). {\sl Middle:} Simulated ratio distribution assuming a random-walk in frequency and variable count rate (see text). The gray underlying histogram in the first two panels shows the actual observed ratio distribution of twin QPO peaks \citep[the data in both panels are those discussed in][]{tor-etal:2007}.
{\sl Right:} The individual distributions of lower and upper QPO frequencies from the random walk simulation. Black-shaded portions of bars represent simultaneous occurrences of both modes (twin QPOs) as shown in the middle panel.}
\label{fig:simulation}
\end{figure*}


\section{Random walk distribution model}

As previously noticed in several works and further suggested by \cite{bel-etal:2005}, the observed time evolution of QPO frequency appears consistent with a series of random walks.

This has been later critised by \citet{bulik:2005} who pointed out that contrary to the distributions of QPOs that appear qualitativelly similar at different times, distributions arising from random walk differ significantly among different realisations (with different seeds). Nevertheless, using a simple model of random walk we attempt to at least roughly reproduce the frequency ratio distribution.

Starting with $\nuL\!=\!700$Hz, we model a long-term evolution of QPO frequencies over 10000 consequent segments. Each segment consists of 50 steps, where one step is assumed to represent 32 seconds of a real observation. An independent random variation of $\pm2\Hz$ in $\nuL$ is assigned to each step. This setup roughly corresponds to the documented frequency drifting through $32$~sec integration intervals \citep{bar-etal:2004,pal-etal:2004} and each segment then mimic 1.6 kiloseconds of QPO evolution.

The QPO frequencies are averaged over each segment, and the linear correlation $\nuU=0.701\nuL+520\Hz$ is considered, so that finally we obtain 10000 frequency pairs. To start with, we assume constant observational conditions (i.e.\ count rate and observing time), adjusting \mbox{$k\!=\!1$}. For each point we calculate its significance based on observed profiles of {\sl Q} and {\sl rms}, which are based merely on datapoints corresponding to twin peak QPO observations. Only such points are considered in the simulation, where both upper and lower QPOs have significance above $3\sigma$ level. The resulting histogram of frequency ratios shows strong clustering around 3:2 ratio, however, it does not reproduce the second peak around $5/4$, which indicates that the assumption of constant count rate my not be sufficient.

As a second step, we adopt an additional (still very simplifying) assumption to the simulation that countrate is varying with frequency. The motivation here comes from a known fact that for a given source there is not a global correlation between source luminosity and QPO frequency, but the two quantities stay correlated during individual (temporary) observational events \citep[so-called parallel-track phenomenon, e.g.,][]{men-etal:1999}. In the case of \4u1636, the maximal countrates related to the highest observed lower QPO frequencies (up to $950\Hz$) are 2--3 times higher than the highest countrates at $\nuL\!\sim\!500$--$700\Hz$ \citep[see Figure 2 in ][]{bar-etal:2005b}. Thus, we keep countrate constant up to \mbox{$\nuL\!\sim\!700\Hz$} and then it is linearly increased with frequency, being about 2.5 times higher at \mbox{$\nuL\!\sim\!950\Hz$} than at \mbox{$\sim\!500$--$700\Hz$}.

In the middle panel of Figure~\ref{fig:simulation}, we show first the histogram of simultaneous occurrences of both QPO modes from our simulation on the background of the observed distribution, and in the right panel of the same figure we plot also individual simulated distributions of lower and upper QPOs. Focusing on twin QPO occurrences, we have a broad peak around $3/2$ and also we obtain a more narrow peak near $5/4$. While the presence of the $3/2$ clustering seem to be very solid and can be reproduced with any setup, the second $5/4$ peak is more subtle feature and depends much on assumed behaviour of countrate. In a real observation, its presence would apparently rely on actual source conditions (and how they would change during the observation) as well as on how the consequent analysis is done. For instance the data examined in \cite{bel-etal:2007b} does not exhibit QPO detections  above $\nuU\!\sim\!1000\Hz$ while the data used in \cite{tor-etal:2007} does. Similarly, if we put e.g.\ more stiff limit on significance or consider lower countrates, we would loose the $5/4$ peak.


\section{Conclusions}

Focused on the atoll source 4U 1636-53 we demonstrate that at frequencies, where the both QPO modes have comparable properties, there is a high probability of detecting both peaks of a twin pair simultaneously. We have found a precise match comparing the observed twin QPO distribution with our simulation based on the observed correlations between QPO frequencies and their properties. The simulation not only reproduces the observed clustering, but it also shows the \qm{complementarity} between upper and lower QPO distributions that has been noticed by \cite{tor-etal:2007}. This suggests that the ratio clustering may origin in the exchange of
dominance between the two modes when one mode fades in and the other one fades out.

Even if the intrinsic distributions of both the mode frequencies were uniform, there would be a non-trivial profile of the observed distributions and clustering of the twin peak detections around certain points (narrow regions) prominent due to behaviour of the QPO amplitudes and coherence times determined by the QPO mechanism. It will require a further detailed analysis to investigate whether the above influence of the QPO properties can explain the ratio clustering observed in 4U 1636-53 completely. For a further understanding of the ratio clustering mechanism (and importance) it is also highly needed to perfom a similar analysis for the other sources. For instance, a very recent study of the atoll source 4U 1820-30 \citep{bar-buj:2008} found that a point
close to the $4/3$ value, where the ratio distribution clusters in that source, and where the amplitudes and quality factors are comparable, is most likely prominent in the intrinsic distribution\footnote{Note also
that in contrary to the case of 4U 1636 they reported a lack of the twin QPO detections close to the 3/2 value, while the amplitudes and quality factors are comparable there as well.}.
\medskip

{\footnotesize{\bf{Acknowledgements.}}
We thank  M.~M\'{e}ndez for several discussions on the subject and, especially, we are thankful to D.~Barret for ideas, comments and for providing the data and software on which this paper builds. We are grateful to W. Klu{\'z}niak and Tomek Bulik for several comments and advice. We thank the referee for all suggestions. We also thank the Yukawa Institute for Theoretical Physics at Kyoto University, where this work was initiated during the YITP-W-07-14 on "Quasi-Periodic Oscillations and Time Variabilities of Accretion Flows". The authors are supported by the Czech grants MSM~478130590384 and LC06014, and by the Polish grants KBN N203~009~31/1466 and 1P03D~005~30.}


\end{document}